\documentclass[twocolumn,showpacs,preprintnumbers,superscriptaddress]{revtex4}
\usepackage{graphicx}
\usepackage{dcolumn}
\usepackage{bm}
\usepackage{color}
\usepackage{amsmath}
\usepackage{amssymb}
\usepackage{amsfonts}
\usepackage{esint}

\begin{document}

\title{Topological insulator and particle pumping in a one-dimensional shaken optical lattice}
\author{Feng Mei}
\email{tianfengmei@gmail.com} \affiliation{National Laboratory of
Solid State Microstructures, Department of Physics, Nanjing
University, Nanjing, China} \affiliation{Centre for Quantum
Technologies, National University of Singapore, 3 Science Drive 2,
Singapore 117543, Singapore}

\author{Jia-Bin You}
\affiliation{Centre for Quantum
Technologies, National University of Singapore, 3 Science Drive 2,
Singapore 117543, Singapore}

\author{Dan-Wei Zhang}
\affiliation{Laboratory of Quantum Information Technology and SPTE,
South China Normal University, Guangzhou 510006, China}

\author{X. C. Yang}
\affiliation{Laboratory of Quantum Information Technology and SPTE,
South China Normal University, Guangzhou 510006, China}

\author{R. Fazio}
\email{r.fazio@sns.it}
\affiliation{Centre for Quantum Technologies, National University of Singapore,
3 Science Drive 2, Singapore 117543, Singapore}
\affiliation{NEST, Scuola Normale Superiore and Istituto Nanoscienze-CNR, I-56126 Pisa, Italy}

\author{Shi-Liang Zhu}
\email{slzhu@nju.edu.cn} \affiliation{National Laboratory of Solid
State Microstructures, Department of Physics, Nanjing University,
Nanjing, China}

\author{L. C. Kwek}
\email{cqtklc@nus.edu.sg}
\affiliation{Centre for Quantum Technologies, National University of Singapore,
3 Science Drive 2, Singapore 117543, Singapore}
\affiliation{Institute of Advanced Studies, Nanyang Technological University, 60 Nanyang View, Singapore 639673}
\affiliation{National Institute of Education, Nanyang Technological University, 1 Nanyang Walk, Singapore 637616}

\date{\today}

\begin{abstract}
  We propose a simple method to simulate and detect topological insulators with cold atoms trapped in a
  one-dimensional bichromatic optical lattice subjected to a time-periodic modulation. The tight-binding form of this shaken system is equivalent to the periodically driven Aubry-Andre model.
  We demonstrate that this model can be mapped into a two-dimensional Chern insulator model, whose energy spectrum hosts a topological phase within an experimentally accessible parameter regime. By tuning the laser phase adiabatically, such one-dimensional system constitutes a natural platform to realize topological particle pumping. We show that the Chern number characterizing the topological features of this system
  can be measured by detecting the density shift after one cycle of pumping.
\end{abstract}

\maketitle

\section{Introduction}
Since the discovery of  topological insulators, the search for
topological state of matter has attracted intense interests in
past years in condensed-matter physics \cite{TIREV1,TIREV2} and atomic, molecular, and optical physics \cite{GFREV}. One
of the recent theoretical advances in this field is the
introduction of Floquet topological insulators
\cite{FTI1,FTI2,FTI3,FTI4,FTI5,FTI6,FTI7}. It is shown that a
topological trivial system can become nontrivial in the presence
of time-periodic perturbations. The classification of Floquet
topological insulator is based on analyzing the
properties of the Floquet operator \cite{FTIC1,FTIC2}. In
particular, these periodic perturbations could be easily achieved by
externally shining laser fields on the system with microwave frequencies
\cite{FTI1,FTI2,FTI3,FTI4,FTI5,FTI6,FTI7}. However, in solid state
system, the materials with topological features are still
quite scarce and the tools to engineer the topological phase
remain limited.

Ultracold atoms trapped in an optical  lattice nowadays have been
widely recognized as powerful tools to simulate and study
many-body problems originally from condensed matter physics
\cite{QSREV1,QSREV2}. This arises from the fact that the cold
atomic systems can provide a clean platform without lattice disorder and even some extreme physical situations unachieved
in condensed matter physics can be reached here. Experimentally,
great progress has been achieved recently in realizing
artificial  gauge fields \cite{GF1,GF2,GF3,GF4}, spin-orbit
coupling for ultracold atoms \cite{SOC1,SOC2,SOC3} and spin Hall
effects \cite{Zhu2006}, which turn cold atoms into a new platform
for simulating topological phases. Some proposals have been put
forward to use this system to mimic quantum anomalous Hall
insulator (Chern insulator) \cite{CI1,CI2,CI3,CI4,CI5},
time-reversal invariant topological insulator \cite{TRI1,TRI2},
and Majorana fermions \cite{Zhu2011}. In particular, cold atoms in modulated optical lattices constitutes a versatile platform to realize synthetic gauge fields and topological phases \cite{FTIOL}. For instance, such setups have been considered recently to experimentally realize the Haldane \cite{Haldane,CI1,CI3} and the Hofstadter model \cite{Hofstadter}, using the technology of modulated optical lattices   \cite{expTI,expTI2}. These experiments led to the first experimental determination of the Chern number using cold atoms \cite{expTI2}.

Remarkably, recent studies have also  shown that some topological
properties of the two-dimensional (2D) integer quantum Hall
insulator could be simulated with one-dimensional (1D) Aubry-Andre
(AA) model \cite{AA}. When the 1D model depends on a
parameter in a periodic manner, one can regard this system as a 2D system, where the physical dimension is extended by the 1D parameter space. Then the topological
property of this system can be characterized by the Chern number
defined on a torus formed by a spatial dimension as well as a
parameter dimension. For cold atomic system, the AA model has been
experimentally realized in a 1D optical lattice for studying
Anderson localization \cite{AAOL} and recently been discussed for
simulating quantum Hall insulator \cite{chen1,mei} and topological
bosonic Mott insulator \cite{zhu,chen2} and charge pumping \cite{Marra}. This idea has also been
generalized to realize the Haldane insulator with a 1D extended
Su-Schrieffer-Heeger model \cite{chen3}.  Compared with the
previous methods for realizing topological phase, the present method does
not need the engineering of synthetic gauge fields, which provides an alternative simple route
to probe the topological feature of quantum Hall states. Another route is also offered by the so-called synthetic dimensions, which use the internal states of the atoms \cite{syn_dim}.

In this paper, we take one step further and show that the
AA model can have the same topological feature as
the 2D Chern insulator, when it is subjected to periodic driven. The 2D Chern insulator is different from the standard quantum Hall insulator,
as it does not require a magnetic field to break time reversal symmetry and thus, it does not display Landau levels. Its discovery stimulated
the search for different exotic topological phases, including $Z_2$ topological insulator and topological superconductor \cite{TIREV1,TIREV2}. In this paper,
we focus on its realization using cold atoms trapped in a 1D shaken bichromatic optical
lattice, but we note that it can also be achieved in 1D quasicrystal systems \cite{PFTI}. When the lattice modulation is in the high frequency regime, we find that the energy spectrum associated with the effective time-independent Hamiltonian displays a gapped Chern
insulator phase. In contrast to the recent proposals for realizing
Floquet topological insulator in 2D optical lattice \cite{FTIOL},
one of the advantages of this 1D framework is that topological pumping arises naturally. This pumping is
topologically protected because the 1D model shares the same
topological origin as the 2D Chern insulator. The number of
pumped particles can be expressed as the Chern number of this
system. Then we employ such pumping to detect the Chern number, hence
characterizing the topological property of the system.

The paper is organized as follows: Section II introduces the
periodically driven AA model, which can be realized with
ultracold atoms trapped in a shaken bichromatic optical
lattice. Section II presents the effective Hamiltonian of the
periodic driven AA model. In section III, we study some
topological features of the system, such as Chern number and edge
states. In section V, we present a feasible approach to detect the
topological Chern number based on topological pumping. A short
conclusion is given in Sec. VI.

\section{Periodically driven AA model}

We consider ultracold fermionic atoms trapped  in a shaken
bichromatic optical lattice. This lattice is generated by the
superposition of two shaken optical lattices. The single particle
Hamiltonian of an atom in this shaken lattice system is written as
\begin{equation}
\label{Hs}
H_s=\frac{p^2_x}{2M}+V_1\sin^2[k_1(x-x_1(t)]+V_2\sin^2[k_2(x-x_2(t)+\phi/2],
\end{equation}
where $V_i$, $k_i=2\pi/\lambda_i$ and $\lambda_i$ ($i=1,2$) are the lattice depth and laser wave vector and wave length. $x_i(t)=b_i\sin(\omega t)$ is the periodic time-dependent lattice shaken. $\omega$ is lattice shaken frequency and $\phi$ is the phase of the second laser. Experimentally, a shaking sinusoidal lattice can be realized through a modulation of the driving frequency and by changing the relative phase of the acousto-optic modulators \cite{chin}. Here we assume the two lattices experience the same shaking amplitude $b=0.2$ which is within the current experimental technology \cite{chin}, and describe the two lattice shaking amouns as $b_1=b\lambda_1/2$ and $b_2=b\lambda_2/2$. In the following, we choose the lattice spacing $a=\lambda_1/2=1$ and $\hbar=1$, and assume the first lattice depth $V_1$ is much bigger than the second lattice depth $V_2$. In the absence of lattice shaking $b=0$, the tight-binding Hamiltonian from
Eq.(\ref{Hs}) leads to the so-called
AA model, which is widely used in the investigation of Anderson
localization as well as some topological phases in cold atomic
physics \cite{AAOL,chen1,mei,zhu,chen2}. In this paper, we refer the
tight-binding model from Eq.(\ref{Hs}) with $\theta_i(t) \propto \sin\omega t$ as
periodically driven AA model. Considering a unitary rotation, the Hamiltonian is
transferred to a new frame $x\rightarrow x+b_1\sin(\omega t)$, and it displays a shaking-induced vector potential
\begin{eqnarray}
H_{r}&=&H_1+H_2,  \\
H_1&=&\frac{(p_x-A_x)^2}{2M}+V_1\sin^2(k_1x), \nonumber\\
H_2&=&V_2\sin^2(k_2x-A_{\phi}+\phi/2), \nonumber
\end{eqnarray}
where the induced vector potentials are
$A_x=\alpha_1\cos(\omega t)$ and $A_{\phi}=\alpha_2\sin(\omega
t)$, with the amplitudes $\alpha_1=M\omega b_1$ and $\alpha_2=k_2(b_2-b_1)$.

We assume all the atoms are trapped in the lowest  band of the
optical lattice, then the Hamiltonian in the second quantization
formalism is of the form
\begin{equation}
H=\int dx \Psi^{\dag}(x)H_r\Psi(x).
\end{equation}
Expanding the field operator in terms of
the Wannier functions $\Psi(x)=\sum_{n}c_nw(x-x_n)$, one can omit
the constant energy terms and get the tight-binding Hamiltonian
\begin{equation}
H=\sum_nJ(t)(c^{\dag}_{n+1}c_n+h.c.)-\sum_{n}\Delta\cos(2\pi\beta n-A_{\phi}+\phi)c^{\dag}_nc_n,\\
\end{equation}
where $\beta=k_2/k_1$, $J(t)=\int dx w(x-x_{n+1})H_1w(x-x_n)$,
$\Delta=\frac{V_2}{2}\int dx w(x)\cos(2k_2x)w(x)$ and the driving amplitude in $A_{\phi}$ has been modified as $\alpha_2=2k_2(b_2-b_1)$.  Note that this
tight-binding model is the Aubry-Andre model, but with an additional periodic
driven contained in the induced gauge potentials $A_{x,\phi}$. Here only the on-site contribution of the second optical
lattice is kept because we assume $V_1$ is much bigger than $V_2$.

In order to realize a two band Chern insulator, we choose
$k_1=2k_2$, then $\beta=1/2$. In
this case, the odd and even number lattice sites feel different
on-site energies. We label this two sites as $a$ and $b$ and use
them to constitute a pseudospin. By employing Fourier
transformation and Peierls substitution, the above Hamiltonian can
be rewritten in the momentum space as $H=\sum_kC^{\dag}_k\mathcal
{H}(k)C_{k}$, where $C_k=(a_k,b_k)^{T}$. The Hamiltonian density
has the form
\begin{equation}
\begin{aligned}
\mathcal{H}(k)=2t_x\cos(k_x-A_x)\sigma_x+\Delta\cos(\phi-A_{\phi})\sigma_z,
\end{aligned} \label{Ht}
\end{equation}
where $t_x$ is the bare hopping without time-dependent modulation, $\sigma_{x,y,z}$ are the Pauli matrices spanned by $a_k$ and
$b_k$. In the Gaussian approximation for the Wannier function of the
ground state, the hopping rate and the on-site energy can be
derived as $t_x=(4/\sqrt{\pi})V_1^{3/4}E^{1/4}_{R1}\text{exp}(-2\sqrt{V_1/E_{R1}})$
and $\Delta=(V_2/2)\text{exp}(-\beta^2/\sqrt{V_1/E_{R1}})$, where
the recoil energies $E_{Ri}=\hbar k_i^2/2M$ ($i=1,2$). Interestingly, by associating the laser-induced potential $A_{\phi}$  with the vector potential $A_y$,
the on-site energies $\Delta$ with twice the hopping rate $2t_y$,
and the laser phase $\phi$ with the quasimomentum $k_y$, one finds that the above one dimensional driving model can be mapped
into the two-dimensional periodically driven $\pi$-flux Harper
model. In the following, we will show that the original gapless quasienergy spectrum will be driven into
the analog of a two-dimensional gapped topological phase.

\section{The effective Hamiltonian}
The essential feature of the above time-dependent Hamiltonian can be captured by an effective time-independent Hamiltonian. This strategy has been extensively studied recently in periodically driven systems \cite{EF,FH,Goldman_Floquet}. Before using this method to derive the effective Hamiltonian of Eq. (\ref{Ht}), we expand the Hamiltonian density $\mathcal {H}(k)$ as
\begin{equation}
\mathcal{H}=\mathcal {H}_0+\sum^{\propto}_{m=1}(\mathcal {H}_me^{im\omega t}+\mathcal {H}_{-m}e^{-im\omega t}) \label{Hk},
\end{equation}
where each expanding component $\mathcal {H}_m$ has been modified by the Bessel functions $J_m(\alpha_1,\alpha_2)$ (see the appendix A). Based on the above expanding Hamiltonian, following the strategy in \cite{Goldman_Floquet}, the effective Hamiltonian of the above equation can be expressed as \cite{footnote}
\begin{equation}
\mathcal {H}_{eff}=\mathcal {H}_0+\frac{1}{\omega}\sum_{m=1}\frac{1}{m}[\mathcal {H}_m,\mathcal {H}_{-m}], \label{Heff}
\end{equation}
where we have omitted high orders of $1/\omega$ \cite{Goldman_Floquet}. When a high driving frequency ($\omega=8t_x$) is employed, the contribution of high orders can be safely ignored. In Fig. 1(a-b), we have numerically demonstrated this point through calculating the modification of the second order terms on the bulk and edge energy spectrums. The results show that, the second order terms only have small influence on the envelope of the spectrum and will not change the gap and the edge spectrum.

In the practical experiment, a weak lattice shaken ($b<1$) is preferred so that the induced heating on the lattice is small. If the lattice shaken is tuned so that the vector potential amplitudes $\alpha_{1,2}<1$, one can find that $J_m(\alpha_{1,2})\sim10^{-m} (m\geq3)$. In this case, in Eq. (\ref{Hk}), only the one- and two-photon transition dominate the whole dynamics. Then we can write Eq. (\ref{Hk}) as
\begin{equation}
\mathcal {H}(k,t)=\mathcal {H}_0+\sum\mathcal
{H}_{m=\pm1,\pm2}e^{im\omega t},
\end{equation}
where
\begin{eqnarray}
\mathcal {H}_0 &=&2t_xJ_0(\alpha_1)\cos(k_x)\sigma_x+\Delta J_0(\alpha_2)\cos(\phi)\sigma_z,\nonumber\\
\mathcal {H}_1&=&2t_xJ_1(\alpha_1)\sin(k_x)\sigma_x-i\Delta J_1(\alpha_2)\sin(\phi)\sigma_z,\nonumber\\
\mathcal {H}_2&=&-2_xtJ_2(\alpha_1)\cos(k_x)\sigma_x+\Delta J_2(\alpha_2)\cos(\varphi)\sigma_z,\\
\mathcal {H}_{-1}&=&\mathcal {H}^+_1,\nonumber\\
\mathcal {H}_{-2}&=&\mathcal {H}_2, \nonumber \label{H012}
\end{eqnarray}
By substituting the above equation into Eq. (\ref{Heff}), we can obtain an effective two band Chern insulator model Hamiltonian
\begin{equation}
\mathcal {H}_{eff}=h_x\sigma_x+h_y\sigma_y+h_z\sigma_z,
\end{equation}
where
\begin{eqnarray}
h_x&=&2t_xJ_0(\alpha_1)\cos(k_x),\nonumber\\
h_y&=&\frac{8}{\omega}t_x\Delta J_1(\alpha_1)J_1(\alpha_2)\sin(k_x)\sin(\phi),\\
h_z&=&\Delta J_0(\alpha_2)\cos(\phi). \nonumber
\end{eqnarray}
As has been shown before, the above 1D atom-lattice system without
shaking can be described by the 2D $\pi-$flux model.  The
x-direction momentum $k_x$ and the laser phase $\phi$ constitute
the 2D space $(k_x,\phi)$, which form a 2D torus under the
periodic boundary condition along the x-direction. In Fig. 1(a),
we plot the energy spectrum with driving in the first Brillouin zone of this
artificial 2D space $k_x\in[0,\pi]$ and $\phi\in[-\pi,\pi]$. In
the absence of shaking, the energy spectrum is gapless and there are
two Dirac points, $D_1=(\frac{\pi}{2},\frac{\pi}{2})$ and
$D_2=(\frac{\pi}{2},-\frac{\pi}{2})$.
However, when the periodic shaking is applied on the optical
lattice, the system will be driven into a gapped insulator phase
by the induced y component $h_y$ in the Eq. (11).

\begin{figure}[tbp]
\includegraphics[width=9.5cm,height=10cm]{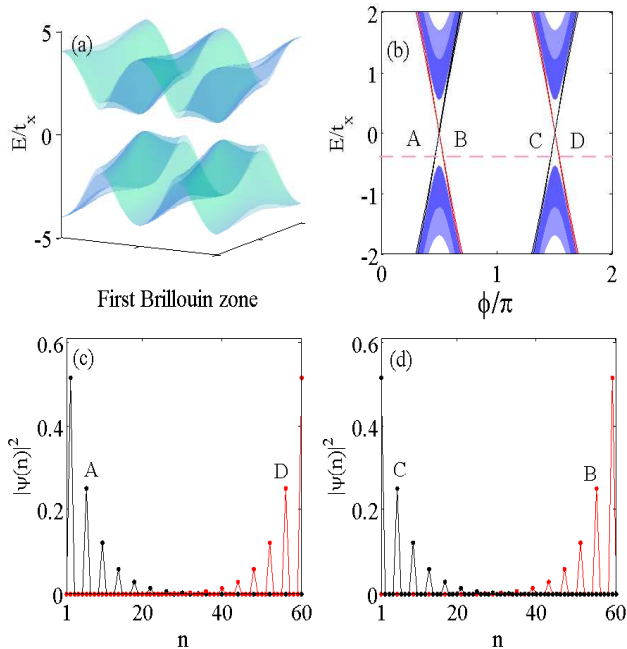}
\caption{(color online) (a) The bulk energy spectrum for driving AA model. (b) The edge state
spectrum when an open boundary condition in the lattice direction is explored. In both cases, the second order corrections for Eq. (\ref{Heff}) have been calculated and plotted in the shaded spectrum. The dashed line in (b) denotes the fermi energy $E_f\simeq-0.4t_x$.  In this case, there are four edge modes travelling in the gap labelled by A, B, C and D. We also have plotted the density distribution of the edge modes A and D in (c) and B and C in (d). The results show that the density of edge modes A and D will occupy the $b$ site of each unit cell and the modes B and C will be in the $a$ site of each unit cell. The lattice length is $L=60$ in the numerical calculation and the other involved parameters are $V_1=6E_{R1}$, $V_2=0.8E_{R1}$, $b=0.2$, $t_x\simeq0.1E_{R1}$, $\omega=8t_x$, $\Delta=6t_x$, $\alpha_1\simeq0.8$ and $\alpha_2\simeq0.6$. }
\end{figure}

\section{Chern insulator and Edge state}

In the following, we will show that this gapped phase is indeed a
topological insulator state by calculating the  Chern number of
the occupied ground band and the edge state spectrum through
exploring the open boundary condition. With a unitary rotation,
$\sigma_x\rightarrow\sigma_y$, $\sigma_z\rightarrow\sigma_x$ and
$\sigma_y\rightarrow\sigma_z$, the above Hamiltonian has
the standard form of the Chern insulator model. This rotation
communicates with time reversal symmetry operator and then will
not change the topological feature of the original model. Based on
a mapping from the Brillouin zone torus $(k_x,\phi)$ to a
spherical surface $S^2$, the Chern number of the occupied ground
band is expressed as
\begin{equation}
C=\frac{1}{4\pi}\int\int
dk_xd\phi(\partial_{k_x}\textbf{h}\times\partial_{\phi}\textbf{h})\cdot\textbf{h},
\end{equation}
where the unit vector field $\textbf{h}=(h_x,h_y,h_z)/h$ with
$h=\sqrt{ h_x^2+h_y^2+h_z^2}$. The Chern number of this two band
model can be derived analytically by calculating the sign of  the
above Jacobian mapping at the two Driac points \cite{usov}
\begin{equation}
C=\frac{1}{2}\sum_{i=1,2}\text{sgn}[(\partial_{k_x}\textbf{h}\times\partial_{\phi}\textbf{h})\cdot\textbf{h}]|_{D_i}.
\end{equation}
After substituting Eq. (11) into the above equation, the Chern
number of this two band model can be obtained as
\begin{equation}
C=-\text{sgn}(J_0(\alpha_1)J_0(\alpha_2))\text{sgn}(J_1(\alpha_1)J_1(\alpha_2)).
\end{equation}
Because the shaking amplitude is small in our system, $\alpha_1$ and
$\alpha_2$ are smaller than one, the rate of Bessel function is
always positive and  the Chern number $C=-1$, thus the system is
in the topological insulator phase. This is quite different from
the original gapless semimetal phase without driving.

The appearance of edge state is another hallmark of topological
state. To show the behavior of edge state, in Fig. 1(b), we choose
the  open boundary condition in the x direction and plot the edge
state of the energy spectrum when the lattice driving is
applied. When a fermi energy is chosen, there are four gapless edge
states travelling in the gap. The velocities of edge modes in the two different edges are opposite.
This point can be seen from the spatial density distribution of the four edge modes in Fig. 1(c-d).
The results also show that the density of edge modes A and D will occupy the $b$ site of each unit cell and the modes B and C will stay in the $a$ site of each unit cell. Furthermore, one can find that, the density of the edge modes in the region $\phi\in(-\pi/2,\pi/2)$ would always occupy the $a$ tybe site in each unit cell, while for other region, the density would occupy the $b$ type site in each unit cell. This arises from the fact that the eigenstate of our 1D lattice model is the eigenstate of pauli matrix $\sigma_z$. The $a$ and $b$ site occupation correspond to the negative and positive eigenstate of $\sigma_z$. In the same way, the bulk states also have this feature. This also can explain the topological particle pumping in the next section. Through tuning the laser phase from $0$ to $2\pi$, the initial density occupying the $b$ site in each unit cell will shifted to the $a$ site of the next unit cell and finally into the $b$ site of this unit cell, which means that the charge will shift by one unit cell.

\begin{figure}
\begin{tabular}{cc}
\hspace{-0.04\textwidth}\includegraphics[width=4cm]{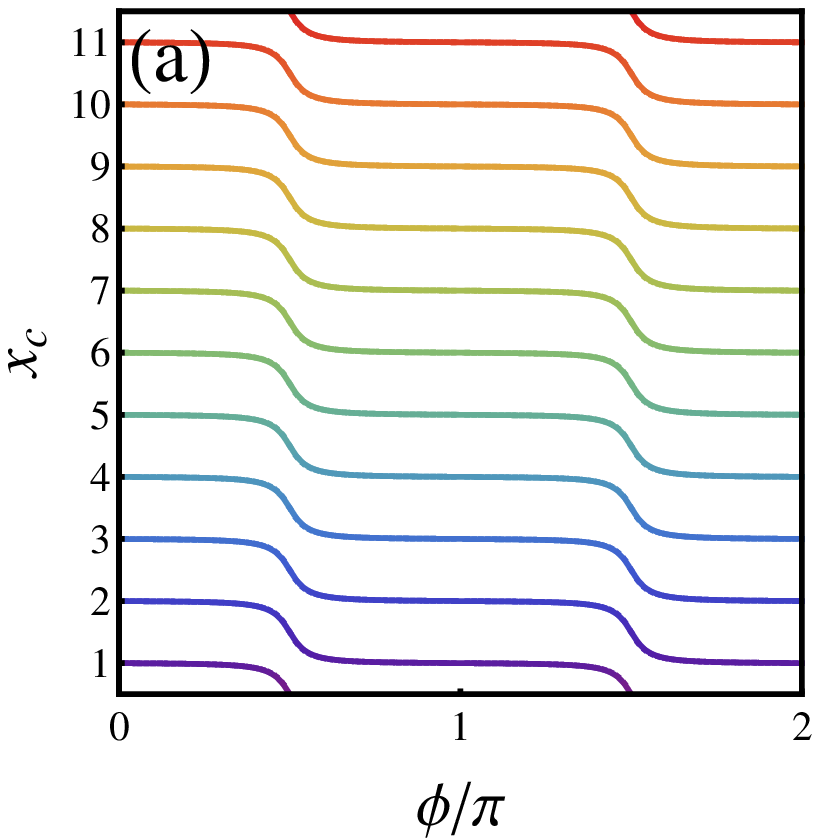} &
\includegraphics[width=4.4cm,height=4.87cm]{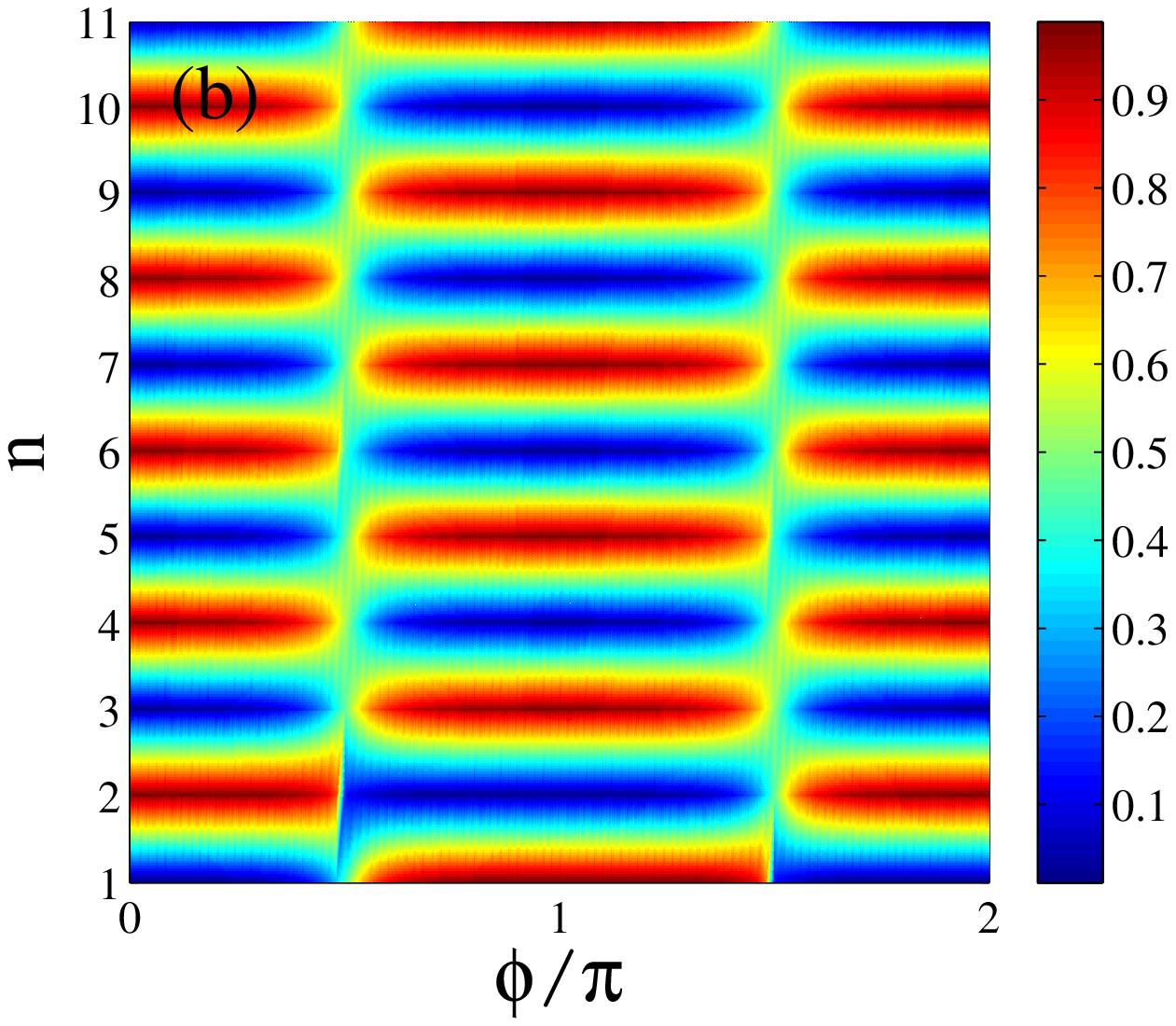}\\
\end{tabular}
\caption{(color online)(a) The Wannier center and (b) The atomic density distribution along the optical
lattice vary with the laser phase $\phi$. After one periodic
pumping by tuning the laser phase over one period, the density
will shift downwards one unit cell. This phenomenon can be observed
in the whole lattice; however, for better illustrating, here we
only plot the density distribution of the lattice in a particular
range $n\in{[1,11]}$. Here the harmonic trap strength
$V_t=0.001$, and other parameters are same
with those in Fig.1.}
\end{figure}

\section{Topological particle pumping}

One of the advantages of the above 1D framework
simulating topological phase is that topological pumping
can be established naturally. Topological pumping in 1D
system was discovered by Thouless \cite{Thouless} and he made the
surprising observation that certain band insulators can provide
quantized charge transport based on an adiabatic pumping. This topological argument also applies for finite system coupled to leads \cite{Brouwer}. The
charge transferred in each pumping cycle is exactly quantized and
can be expressed as the Chern number
\begin{equation}
Q=\frac{1}{2\pi}\int\int dk_xdt\mathcal {F}(k_x,t)=C[k_x,t],
\end{equation}
where $C[k_x,t]$ is the Chern number defined on the time and momentum Brillouin zone space, and $\mathcal{F}(k_x,t)$ is the Berry curvature. Topological pumping has also been discussed recently in the context of cold atoms \cite{zhang,davide,wlei1} and quantum wire systems \cite{marco,berg,loss}.

In the following, we will show the topological pumping can be directly realized based on
the proposed 1D framework here. Because our proposed 1D model has the
same topological feature as the 2D Chern insulator, its ground
state is a topological phase and the corresponding pumping is
topologically protected. By slowly tuning the laser phase over one
period, the number of the pumping particle number is just the Chern number
of the ground band. The number of the pumping particle can be connected with
the Wannier center based on the modern charge polarization
theory \cite{polar}. It states that charge polarization can be related to the Wannier center by
\begin{equation}
P=x_c=\langle\mathcal {W}_n|\hat{x}|\mathcal {W}_n\rangle=\frac{1}{2\pi}\int dk_x A(k_x),
\end{equation}
where $|\mathcal {W}_n\rangle$ is the Wannier function of the ground band associated with the unit cell n, $\hat{x}$ is the position operator and $A(k_x)$ is the corresponding Berry connection. If the Hamiltonian $H$ depends on a parameter, the change of the Wannier center with this parameter is a gauge invariant quantity. In our 1D framework, this parameter is the laser phase. If we tune this phase from $\phi_i=0$ to $\phi_f=2\pi$, $H(\phi_i)=H(\phi_f)$, based on Stokes theorem (see appendix B for details), one can find that the change of the Wannier center for each unit cell is the Chern number of the ground band
\begin{equation}
\begin{split}
\Delta x_c &=x_c(\phi_f)-x_c(\phi_i),\\
&=\frac{1}{2\pi}[\int dk_xA(k_x,\phi_f)-\int dk_xA(k_x,\phi_i),\\
&=C[k_x,\phi],
\end{split} \label{wc}
\end{equation}
In Fig. 2(a), we have numerically calculated the change of the Wannier center of each lattice site. After tuning the laser phase over one period, the Wannier center for each lattice site will shift downwards two sites (one unit cell), then the Wannier center for each unit cell will shift downwards one. According to the Eq. (\ref{wc}), the total pumping particle number is one and the Chern number is $-1$. The sign of the Chern number depends on the shift direction.

Experimentally, the change of Wannier center can be inferred from observing the change of the density distribution along the lattice. The similar method has been proposed recently to detect the Chern number for characterizing integer quantum Hall phase in a 2D optical lattice system
\cite{wlei2}, where a time-of-flight measurement along x direction
combined with in situ detection along the y direction is needed. The atomic density is defined by
\begin{equation}
\rho(\phi,n)=\sum_{E_{oc}\leqslant E_f}|\psi_{oc}(\phi,n)|^2,
\end{equation}
where $E_f$ denotes the Fermi energy, $E_{oc}$ denotes the
occupied state of the fermionic atoms and $\psi_{oc}$ is the
corresponding wave function. In Fig. 2 (b), we have plotted the atomic density
distribution along the lattice with the change of the laser phase in the presence of open boundary condition.
In the experiment with cold atomic gases, an external harmonic potential is always present to trap the atomic cloud. We have taken account into its effect by adding the term $H_t=V_t\sum_n(n-L/2)^2c^{+}_nc_n$ in the numerical
calculation, where $V_t$ is the trap strength and $L$ is the
lattice size. Indeed, the harmonic trap strength in the experiment can be tuned to enough small value, the main results of this paper remain intact in this case. Now we take the lattice site $n=1$ as an example to describe the particle pumping. When the laser phase is tuned from
$\phi=0$ to $\phi=\pi$, the density will change from $\rho=0.01$ to $\rho=0.98$, it means that the density has been shifted downwards (from the lattice site $n>1$ to $n=1$). After tuning the laser phase over one period, the density shifts downwards one unit cell (two lattice sites), which yields the Chern
number is $C=-1$ and the sign dependents on the shift direction. This detection
can be done experimentally by tuning the laser phase slowly and
performing \emph{in situ} density measurements along the lattice
to image the density distribution.

\section{Summary}

In summary, we have shown that the topological features of the
topological insulator can be simulated and probed with a
periodically driven AA model. We realize this 1D model with cold
atoms trapped in a 1D shaken optical superlattice. By introducing
the laser phase as an additional dimension, we have demonstrated
that this 1D model can be mapped into a 2D topological Chern
insulator model. The energy spectrum of the system will open a gap
even for weak driving. Through calculating the Chern number of
the bulk system and the edge states of the system with open
boundary condition, we have shown that this gapped state should be
a topological insulator state. We have also shown that this 1D
framework can form a natural realization of topological
pumping and it allows us to directly measure the Chern number of
the  topological state. This method can be generalized to evanescently coupled helical waveguides for achieving photonic Floquet topological
insulators \cite{PFTI}, where the time periodic modulation is replace by
the spatial modulation and the whole topological pumping process can be
seen directly in the real space. Interestingly, the above dimension reduction method
also provides a new route to simulate the high-dimensional topological phases, including
four dimensional topological insulators. When periodic modulation is applied, it may also allow us
to study non-equilibrium topological phases.

\section{Acknowledgements}

F. Mei thanks N. Goldman for helpful inputs and revisions. S. L. Zhu was supported by the NSFC (Grant No. 11125417), the
SKPBR of China (Grants No. 2011CB922104), the PCSIRT (Grant
No. IRT1243). J. B. You was supported by the NRF of Singapore (Grant No. WBS: R-710-000-008-271). KLC and RF acknowledge support from the National Research Foundation, the Ministry of Education, Singapore.

\appendix
\begin{widetext}
\section{Derivation of Equation (\ref{Hk})}
The periodically driven Hamiltonian density in Eq. (\ref{Ht}) can be expanded as $\mathcal {H}=\sum^{\propto}_{n=-\propto}\mathcal {H}_ne^{in\omega t}$ with the Bessel function. This can be seen by firstly expanding the Hamiltonian as
\begin{equation}
\begin{split}
\mathcal {H}(k)&=2t_x\cos(k_x-A_x)\sigma_x+\Delta\cos(\varphi-A_{\varphi})\sigma_z\\
&=t_x[e^{ik_x}e^{-i\alpha_1\cos(\omega t)}+e^{-ik_x}e^{i\alpha_1\cos(\omega t)}]\sigma_x+\frac{\Delta}{2}[e^{i\varphi}e^{-i\alpha_2\sin(\omega t)}+e^{-i\varphi}e^{i\alpha_2\sin(\omega t)}]\sigma_z. \label{Hexp}
\end{split}
\end{equation}
Using the Bessel function
\begin{equation}
\text{exp}[\alpha\frac{x-x^{-1}}{2}]=\sum^\propto_{m=-\propto}J_m(\alpha)x^m,\\
\end{equation}
we can get
\begin{equation}
\begin{aligned}
\text{exp}[i\alpha_2\sin(\omega t)]=\text{exp}[\alpha_2\frac{e^{i\omega t}-e^{-i\omega t}}{2}]=\sum^\propto_{m=-\propto}J_m(\alpha_2)e^{im\omega t},\\
\text{exp}[i\alpha_1\cos(\omega t)]=\text{exp}[i\alpha_1\sin(\frac{\pi}{2}+\omega t)]=\sum^\propto_{m=-\propto}J_m(\alpha_1)e^{im(\frac{\pi}{2}+\omega t)}.
\end{aligned}
\end{equation}
Through substituting the above equation into Eq. (\ref{Hexp}), one can get
\begin{equation}
\mathcal
{H}=\mathcal {H}_0+\sum^{\propto}_{m=1}(\mathcal {H}_me^{im\omega t}+\mathcal {H}_{-m}e^{-im\omega t}),
\end{equation}
where $H_{-m}=H^{+}_m$. When $m\in$ odd,
\begin{equation}
H_m=x_m\sin{k_x}\sigma_x-iz_m\sin{\varphi}\sigma_z, \label{Hmodd}
\end{equation}
where the coefficients are
\begin{equation}
\begin{split}
x_m&=2\sin(\frac{m\pi}{2})t_xJ_{m}(\alpha_1),\\
z_m&=\Delta J_{m}(\alpha_2),
\end{split}
\end{equation}
While if $m\in$ even,
\begin{equation}
H_m=x_m\cos{k_x}\sigma_x+z_m\cos{\varphi}\sigma_z, \label{Hmeven}
\end{equation}
where the coefficient $x_m$ are changed into
\begin{equation}
x_m=2\cos(\frac{m\pi}{2})t_xJ_{m}(\alpha_1),
\end{equation}
and the expression for $z_m$ is same as the odd case.

\section{Proof of Equation (\ref{wc})}

The Chern number can be written as a surface integral of the Berry curvature which is the curl of the Berry connection $\mathbf{A}(k_{x},\phi)$ over the first Brillouin zone (FBZ). Naively, it can be recast into a line integral of the Berry connection along the boundary of FBZ via Stokes' theorem,
\begin{equation}
\begin{split}
C=\frac{1}{2\pi}\iint_{T^2}dk_{x}d\phi\nabla\times\mathbf{A}(k_{x},\phi)=\frac{1}{2\pi}\ointctrclockwise_{\partial T^2}\mathbf{A}(k_{x},\phi)\cdot\mathbf{dl},\\
\end{split}
\end{equation}
where $T^2$ and $\partial T^2$ are the FBZ and its boundary, respectively. However, since the FBZ is a torus which has no boundary, the Chern number is zero if $\mathbf{A}(k_{x},\phi)$ is well defined in the whole FBZ. Therefore, nonzero values of the Chern number are the consequences of singularity in the FBZ where the Stokes' theorem is invalid.

Let's assume that the occupied state wavefunction $\psi(k_{x},\phi)$ has a singular point $p$ in the entire FBZ. To apply the Stokes' theorem in this case, we can surround this singular point by a closed loop $c_{p}$. On removing this singularity in the wavefunction, we can use different gauges for $\psi(k_{x},\phi)$ inside and outside the loop. Outside the loop, the wavefunction $\psi(k_{x},\phi)$ is well defined; inside the loop, we can do a gauge transform, $\psi\rightarrow e^{if_p}\psi$, to obtain the well-defined wavefunction. Since the Berry curvature is gauge invariant, this transform does not change the value of Chern number. However, the Berry connection is indeed gauge variant. In each region, the Berry connections become
\begin{equation}
\begin{split}
T^2-s_{p}&:\mathbf{A}(k_{x},\phi)=-i\langle\psi|\nabla|\psi\rangle,\\ s_{p}&:\mathbf{A}_{p}(k_{x},\phi)=\mathbf{A}(k_{x},\phi)+\nabla f_{p}(k_{x},\phi),\\
\end{split}
\end{equation}
where $s_{p}$ is the region surrounding by $c_{p}$. Because now the Berry connections $\mathbf{A}$ and $\mathbf{A}_{p}$ are well defined in corresponding regions, we can implement the Stokes' theorem in the following way,
\begin{equation}
\begin{split}
C&=\frac{1}{2\pi}\Big[\iint_{T^2-s_{p}}dk_{x}d\phi\nabla\times\mathbf{A}(k_{x},\phi)+\iint_{s_{p}}dk_{x}d\phi\nabla\times\mathbf{A}_{p}(k_{x},\phi)\Big]\\
&=\frac{1}{2\pi}\Big[\ointctrclockwise_{\partial(T^2-s_{p})}\mathbf{A}(k_{x},\phi)\cdot\mathbf{dl}+\ointctrclockwise_{\partial s_{p}}\mathbf{A}_{p}(k_{x},\phi)\cdot\mathbf{dl}\Big],\\
&=\frac{1}{2\pi}\ointctrclockwise_{c_{p}}[\mathbf{A}_{p}(k_{x},\phi)-\mathbf{A}(k_{x},\phi)]\cdot\mathbf{dl}.\\
\end{split}
\end{equation}
Imagine that $c_{p}$ expands gradually to the boundary of the FBZ, then since $\mathbf{A}(k_{x},\phi)$ is well defined outside the loop $c_p$, the line integral is zero. For the connection $\mathbf{A}_{p}(k_{x},\phi)$, it has a singular point outside the loop. We choose a gauge that the singular point of $\mathbf{A}_{p}(k_{x},\phi)$ locates at the boundary of $\phi=0$ or $\phi=2\pi$, then using the periodicity of $\mathbf{A}_{p}(k_{x},\phi)$, the Chern number can be further reduced to
\begin{equation}
\begin{split}
C&=\frac{1}{2\pi}\ointctrclockwise_{c_{p}}[\mathbf{A}_{p}(k_{x},\phi)-\mathbf{A}(k_{x},\phi)]\cdot\mathbf{dl},\\
&=\frac{1}{2\pi}\ointctrclockwise_{\partial T^2}\mathbf{A}_{p}(k_{x},\phi)\cdot\mathbf{dl},\\
&=\frac{1}{2\pi}\Big[\int_{0}^{2\pi}d\phi\mathbf{A}_{p}(0,\phi)+\int_{2\pi }^{0}d\phi\mathbf{A}_{p}(\pi,\phi)\\
&+\int_{0}^{\pi}dk_{x}\mathbf{A}_{p}(k_{x},2\pi)+\int_{\pi}^{0}dk_{x}\mathbf{A}_{p}(k_{x},0)\Big],\\
&=\frac{1}{2\pi}\Big[\int_{0}^{\pi}dk_{x}\mathbf{A}_{p}(k_{x},2\pi)-\int_{0}^{\pi}dk_{x}\mathbf{A}_{p}(k_{x},0)\Big]\\
&=x_c(\phi_f)-x_c(\phi_i)=\Delta x_c
\end{split}
\end{equation}

\end{widetext}


\begin{thebibliography}{99}

\bibitem{TIREV1} M. Z. Hasan and C. L. Kane, Rev. Mod. Phys. \textbf{82}, 3045 (2010).
\bibitem{TIREV2} X.-L. Qi and S.-C. Zhang, Rev. Mod. Phys. \textbf{83}, 1057 (2011).
\bibitem{GFREV} J. Dalibard, F. Gerbier, G. Juzeli¨±nas, and P. Ohberg, Rev. Mod. Phys. \textbf{83}, 1523 (2011); N. Goldman, G. Juzeliunas, P. Ohberg, and I. B. Spielman, e-print arXiv:1308.6533.
\bibitem{FTI1} N. H. Lindner, G. Refael, and V. Galitski, Nat. Phys. \textbf{7}, 490 (2011); N. H. Lindner, D. L. Bergman, G. Refae, and V. Galitski,
Phys. Rev. B \textbf{87}, 235131 (2013).
\bibitem{FTI2} J.-I. Inoue and A. Tanaka, Phys. Rev. Lett. \textbf{105}, 017401 (2010); M. Ezawa, \emph{ibid}. \textbf{110}, 026603 (2013).
\bibitem{FTI3} B. Dora, J. Caysso, F. Simon, and R. Moessner, Phys. Rev. Lett. \textbf{108}, 056602 (2012).
\bibitem{FTI4} Y. T. Katan and D. Podolsky, Phys. Rev. Lett. \textbf{110}, 016802 (2013).
\bibitem{FTI5} T. Kitagawa, T. Oka, A. Brataas, L. Fu, and E. Demler, Phys. Rev. B \textbf{87}, 235131 (2013).
\bibitem{FTI6} M. Lababidi, I. I. Satija, and Erhai Zhao, Phys. Rev. Lett. \textbf{112}, 026805 (2014).
\bibitem{FTI7} A. G. Grushin, A. Gomez-Leon, and T. Neupert, Phys. Rev. Lett. \textbf{112}, 156801 (2014).
\bibitem{FTIC1} T. Kitagawa, E. Berg, M. Rudner, and E. Demler, Phys. Rev. B \textbf{82}, 235114 (2010).
\bibitem{FTIC2} M. S. Rudner, N. H. Lindner, E. Berg, and M. Levin, Phys. Rev. X \textbf{3}, 031005 (2013).
\bibitem{QSREV1} M. Lewenstein, A. Sanpera, V. Ahufinger, B. Damski, A. Sen De, and U. Sen, Adv. Phys. \textbf{56}, 243 (2007).
\bibitem{QSREV2} I. Bloch, J. Dalibard, and W. Zwerger, Rev. Mod. Phys. \textbf{80} 885
(2008).

\bibitem{GF1} Y.-J. Lin, R. L. Compton, K. Jimenez-Garcia, J.V. Porto, and I. B. Spielman, Nature (London) \textbf{462}, 628 (2009);
K. Jimez-Garcia, L. J. LeBlanc, R. A. Williams, M. C. Beeler, A.
R. Perry, and I. B. Spielman, Phys. Rev. Lett. \textbf{108},
225303 (2012).

\bibitem{GF2} M. Aidelsburger, M. Atala, S. Nascimbene, S. Trotzky, Y.-A. Chen, and I. Bloch, Phys. Rev. Lett. \textbf{107}, 255301 (2011);
M. Aidelsburger, M. Atala, M. Lohse, J. T. Barreiro, B. Paredes, and I. Bloch, \emph{ibid}. \textbf{111}, 185301 (2013).

\bibitem{GF3} J. Struck, C.Olschlager, R. Le Targat, P. Soltan-Panahi, A. Eckardt, M. Lewenstein, P. Windpassinger, and K. Sengstock, Science \textbf{333}, 996 (2011);
 J. Struck, C. Olschlager, M. Weinberg, P. Hauke, J. Simonet, A. Eckardt, M. Lewenstein, K. Sengstock, and P. Windpassinger, Phys. Rev. Lett. \textbf{108}, 225304 (2012).

\bibitem{GF4} H. Miyake, G. A. Siviloglou, C. J. Kennedy, W. Cody Burton, and W. Ketterle, Phys. Rev. Lett. \textbf{111}, 185302 (2013).

\bibitem{SOC1} Y.-J. Lin, R. L. Compton, K. Jimenez-Garcia, W. D. Phillips, J.V. Porto, and I. B. Spielman, Nature Phys. \textbf{7}, 531 (2011).
\bibitem{SOC2} P. J. Wang, Z. Q. Yu, Z. K. Fu, J. Miao, L. H. Huang, S. J. Chai, H. Zhai, and J. Zhang,  Phys. Rev. Lett. \textbf{109}, 095301 (2012);
L. W. Cheuk, A. T. Sommer, Z. Hadzibabic, T. Yefsah, W. S. Bakr, and M. W. Zwierlein, \emph{ibid}. \textbf{109}, 095302 (2012).

\bibitem{SOC3} J. Y. Zhang, S. C. Ji, Z. Chen, L. Zhang, Z. D. Du, B. Yan, G. S. Pan, B. Zhao, Y. J. Deng, H. Zhai, S. Chen, and J. W. Pan,
Phys. Rev. Lett. \textbf{109}, 115301 (2012).

\bibitem{Zhu2006}  M. C. Beeler, R. A. Williams, K. Jimenez-Garcia, L. J.
LeBlanc, A. R. Perry, and I. B. Spielman, Nature (London) 498, 201
(2013); S. L. Zhu, H. Fu, C.-J. Wu, S.-C. Zhang, and L.-M. Duan,
Phys. Rev. Lett. \textbf{97}, 240401 (2006); X. J. Liu, X. Liu, L.
C. Kwek, and C. H. Oh, \emph{ibid}. \textbf{98}, 026602 (2007).

\bibitem{CI1}
L. B. Shao, S. L. Zhu, L. Sheng, D. Y. Xing, Z. D. Wang,  Phys.
Rev. Lett.  \textbf{101}, 246810 (2008); C. Wu, \emph{ibid}.
\textbf{101}, 186807 (2008); D. W. Zhang, Z. D. Wang, and S. L.
Zhu, Front. Phys. \textbf{7}, 31 (2012); Feng Mei, D. W. Zhang, and S. L. Zhu, Chin. Phys. B, \textbf{11}, 116106 (2013).

\bibitem{CI2} T. D. Stanescu, V. Galitski, J. Y. Vaishnav, C. W. Clark, and S. DasSarma, Phys. Rev. A \textbf{79}, 053639 (2009);
T. D. Stanescu,V. Galitski, and S. DasSarma, \emph{ibid}.
\textbf{82}, 013608 (2010).

\bibitem{CI3} E. Alba, X. Fernandez-Gonzalvo, J. Mur-Petit, J. K. Pachos, and J. J. Garcia-Ripoll, Phys. Rev. Lett. \textbf{107}, 235301 (2011); N Goldman, E Anisimovas, F Gerbier, P Ohberg, I. B. Spielman, and G Juzeli¨±nas, New J. Phys. \textbf{15}, 013025 (2013).

\bibitem{CI4} N. R. Cooper, Phys. Rev. Lett. \textbf{106}, 175301 (2011).

\bibitem{CI5} X. J. Liu, X. Liu, C. Wu, and J. Sinova, Phys. Rev. A \textbf{81}, 033622 (2010); X. J. Liu, K. T. Law, and T. K. Ng, Phys. Rev. Lett. 112, 086401 (2014).

 \bibitem{TRI1}  N. Goldman, I. Satija, P. Nikolic, A. Bermudez, M. A. Martin-Delgado, M. Lewenstein, and I. B. Spielman, Phys. Rev. Lett. \textbf{105}, 255302 (2010).
 \bibitem{TRI2} B. Beri and N. Cooper, Phys. Rev. Lett. \textbf{107}, 145301 (2011).

 \bibitem{Zhu2011} S. L. Zhu, L.-B. Shao, Z. D. Wang, and L.-M.
 Duan, Phys. Rev. Lett. \textbf{106}, 100404 (2011); S. Tewari, S. Das Sarma, C. Nayak, C. Zhang, and P.
Zoller, \emph{ibid}. \textbf{98}, 010506 (2007).

\bibitem{FTIOL}  P. Hauke \emph{et al}., Phys. Rev. Lett. \textbf{109}, 145301 (2012); G. C. Liu, N. N. Hao, S. L. Zhu, W. M. Liu, Phys. Rev. A \textbf{86}, 013639 (2012); S. K. Baur \emph{et al}., \emph{ibid}. \textbf{89}, 051605 (2014); Wei Zheng and Hui Zhai, \emph{ibid}. \textbf{89}, 061603 (2014);
S.-L. Zhang and Q. Zhou, e-print arXiv: 1403.0210v1.


 \bibitem{Haldane} F. D. M. Haldane, Phys. Rev. Lett. \textbf{61}, 2015 (1988).

\bibitem{Hofstadter} D.~R. Hofstadter, Phys. Rev. B \textbf{14}, 2239 (1976).



 \bibitem{expTI} G. Jotzu \emph{et al}., Nature (London) \textbf{515}, 237 (2014);

 \bibitem{expTI2} M. Aidelsburger \emph{et al}., e-print arxiv: 1407.4205v1;

 \bibitem{AA} Y. E. Kraus, Y. Lahini, Z. Ringel, M. Verbin, and O. Zilberberg, Phys. Rev. Lett. \textbf{109}, 106402 (2012);
 Y. E. Kraus and O. Zilberberg, \emph{ibid}. \textbf{109}, 116404 (2012); M. Verbin, O. Zilberberg, Y. E. Kraus, Y. Lahini, and
 Y. Silberberg, \emph{ibid}. \textbf{110}, 076403 (2013).

 \bibitem{AAOL} G. Roati \emph{et al}., Nature (London) \textbf{453}, 895 (2008); B. Deissler et al., Nat. Phys. \textbf{6}, 354 (2010); E. Lucioni, B. Deissler, L. Tanzi, G. Roati, M. Zaccanti, M. Modugno, M. Larcher, G. Dalfovo, M. Inguscio, and G. Modugno, Phys. Rev. Lett. \textbf{106}, 230403 (2011).
 \bibitem{chen1} L. J. Lang, X. M. Cai, and S. Chen, Phys. Rev. Lett. \textbf{108}, 220401 (2012).

 \bibitem{mei} F. Mei, S. L. Zhu, Z. M. Zhang, C. H. Oh, and N. Goldman, Phys. Rev. A \textbf{85}, 013638 (2012).
 \bibitem{zhu} S. L. Zhu, Z.-D. Wang, Y.-H. Chan, and L.-M. Duan, Phys. Rev. Lett. \textbf{110}, 075303 (2013).

 \bibitem{chen2} Z. H. Xu and S. Chen, Phys. Rev. B \textbf{88}, 045110 (2013).
 \bibitem{Marra} P. Marra, R. Citro, and C. Ortix, e-print arXiv: 1408.4457v1.

 \bibitem{chen3} L. Li, Z.H Xu, and S. Chen, Phys. Rev. B \textbf{89}, 085111 (2014).
 \bibitem{syn_dim} A. Celi \emph{et al.}, Phys. Rev. Lett. \textbf{112}, 043001 (2014).

 \bibitem{PFTI} M. C. Rechtsman1, J. M. Zeuner, Y. Plotnik, Y. Lumer, D. Podolsky, F. Dreisow, S. Nolte,
M. Segev, and A. Szameit, Nature (London) \textbf{496}, 196 (2013).


\bibitem{chin} C. V. Parker, L.-C. Ha and C. Chin, Nat. Phys. \textbf{9}, 769 (2013).


\bibitem{FH} A. Gomez-Leon and G. Platero, Phys. Rev. Lett. \textbf{110}, 200403 (2013); P. Delplace, A. Gomez-Leon, and G. Platero, Phys. Rev. B \textbf{88}, 245422 (2013); \textbf{89}, 205408 (2014).
\bibitem{EF} S. Rahav, I. Gilary, and S. Fishman, Phys. Rev. A \textbf{68}, 013820 (2003).
\bibitem{Goldman_Floquet} N. Goldman and J. Dalibard, Phys. Rev. X \textbf{4}, 031027 (2014).

\bibitem{footnote} The effective Hamiltonian is equal to the logarithm of the time evolution operator over one driving period, which can be derived through Magnus expansion of the time-ordering evolution operator and making time average of the time-dependent Hamitonian over one driving period. However, the initial time in the time average will enter the final effective Hamiltonian. With the method in \cite{EF,Goldman_Floquet}, this dependence can be eliminated through employing two initial and final time-dependent operators before and after the evolution of the effective Hamiltonian. These kick operations are time-periodic and average to zero over one driving period.

\bibitem{usov} N.A. Usov, Solid State Commun. \textbf{68}, 943 (1988).
\bibitem{Thouless} D. J. Thouless, Phys. Rev. B \textbf{27}, 6083 (1983).
\bibitem{Brouwer} P. W. Brouwer, Phys. Rev. B \textbf{58}, R10135 (1998); M. Buttiker, H. Thomas, and A. Pretre, Z. Phys. B \textbf{94}, 133
(1994).
\bibitem{zhang} Y. Qian, M. Gong, and C. Zhang, Phys. Rev. A \textbf{84}, 013608 (2011).
\bibitem{davide} D. Rossini, M. Gibertini, V. Giovannetti, R. Fazio, Phys. Rev. B \textbf{87}, 085131 (2013).

\bibitem{wlei1} L. Wang, M. Troyer, and X. Dai, Phys. Rev. Lett. \textbf{111}, 026802 (2013).
\bibitem{marco} M. Gibertini, R. Fazio, M. Polini, and F. Taddei, Phys. Rev. B \textbf{88}, 140508(R) (2013).
\bibitem{berg} A. Keselman, L. Fu, A. Stern, E. Berg, Phys. Rev. Lett. \textbf{111}, 116402 (2013).
\bibitem{loss} A. Saha, D. Rainis, R. P. Tiwari, and D. Loss, Phys. Rev. B \textbf{90}, 035422 (2014)
\bibitem{polar} R. D. King-Smith and D. Vanderbilt, Phys. Rev. B \textbf{47}, 1651 (1993).
\bibitem{wlei2} L. Wang, A. A. Soluyanov, and M. Troyer, Phys. Rev. Lett. \textbf{110}, 166802 (2013).













\end{thebibliography}
\end{document}